\documentclass[acmsmall]{acmart}

\usepackage{booktabs} 
\usepackage{algorithm}
\usepackage{algpseudocode}

\setcopyright{acmcopyright}
\acmJournal{PACMCGIT}
\acmYear{2018} 
\acmVolume{1} 
\acmNumber{1} 
\acmArticle{14} 
\acmMonth{5} 
\acmPrice{15.00}
\acmDOI{10.1145/3203192}
\setcopyright{usgovmixed}

\received{November 2017}
\received{March 2018}
\received[accepted]{June 2018}

\begin{document}

\title{Deep Surface Light Fields}

\author{Anpei Chen}
\author{Minye Wu}
\author{Yingliang Zhang}
\affiliation{%
  \institution{ShanghaiTech University}
  \streetaddress{393 Huaxia middle Rd}
  \city{Shanghai}
  \state{SH}
  \postcode{201210}
  \country{China}}
\email{chenap@, wumy@, yingliang@, shanghaitech.edu.cn}

\author{Nianyi Li}
\affiliation{%
  \institution{Duke University}
  \country{USA}
}
\email{nianyi.li@duck.edu}

\author{Jie Lu}
\author{Shenghua Gao}
\author{Jingyi Yu}

\affiliation{%
  \institution{ShanghaiTech University}
  \streetaddress{393 Huaxia middle Rd}
  \city{Shanghai}
  \state{SH}
  \postcode{201210}
  \country{China}}
\email{Lujie1@, Gaoshh@, yujingyi@, shanghaitech.edu.cn}

\begin{abstract}
A surface light field represents the radiance of rays originating from any points on the surface in any directions. Traditional approaches require ultra-dense sampling to ensure the rendering quality. In this paper, we present a novel neural network based technique called deep surface light field or DSLF to use only moderate sampling for high fidelity rendering. DSLF automatically fills in the missing data by leveraging different sampling patterns across the vertices and at the same time eliminates redundancies due to the network's prediction capability. For real data, we address the image registration problem as well as conduct texture-aware remeshing for aligning texture edges with vertices to avoid blurring. Comprehensive experiments show that DSLF can further achieve high data compression ratio while facilitating real-time rendering on the GPU. 
\end{abstract}

%
%


%
%



\maketitle

\renewcommand{\shortauthors}{A Chen et al.}

\section{Introduction}
Reproducing photorealistic appearance of real objects is a core problem in computer graphics. Applications are numerous, ranging from surface reflectance modeling to photorealistic rendering, and to virtual and augmented reality. Most existing approaches fall into two categories: physically based rendering (PBR) and image-based modeling and rendering (IBMR). PBR attempts to faithfully model how light physically transports, e.g., in terms of the radiance transfer functions~\cite{Jannson1980Radiance}. It can produce ultra-high realism at the cost of long computational time. IBMR, in contrast, directly employs images captured under real settings and renders new views from the captured ones. IBMR can achieve real-time performance but its quality depends heavily on sampling density. 

\begin{figure}[t]
\begin{center}
   \includegraphics[width=1\linewidth]{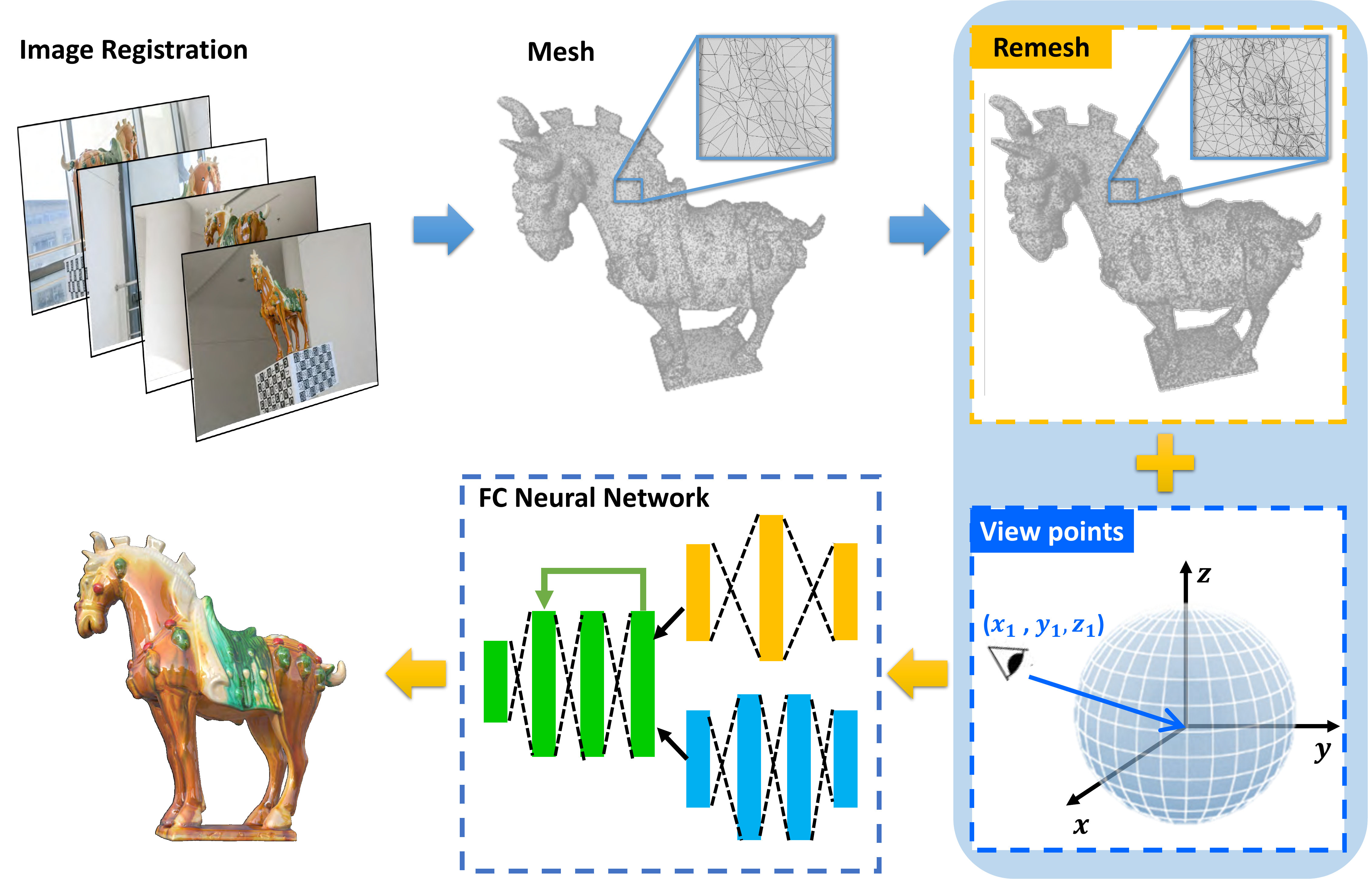}
\end{center}
   \caption{The processing pipeline of our deep surface light field framework: we first obtain the 3D model of the object, then register the input images and conduct texture-aware remeshing, and finally transform the data to feed into the deep network for training. The trained result can be directly used for real-time rendering.}
\label{fig:pipeline}
\end{figure}

\begin{figure}[t]
\begin{center}
   \includegraphics[width=0.9\linewidth]{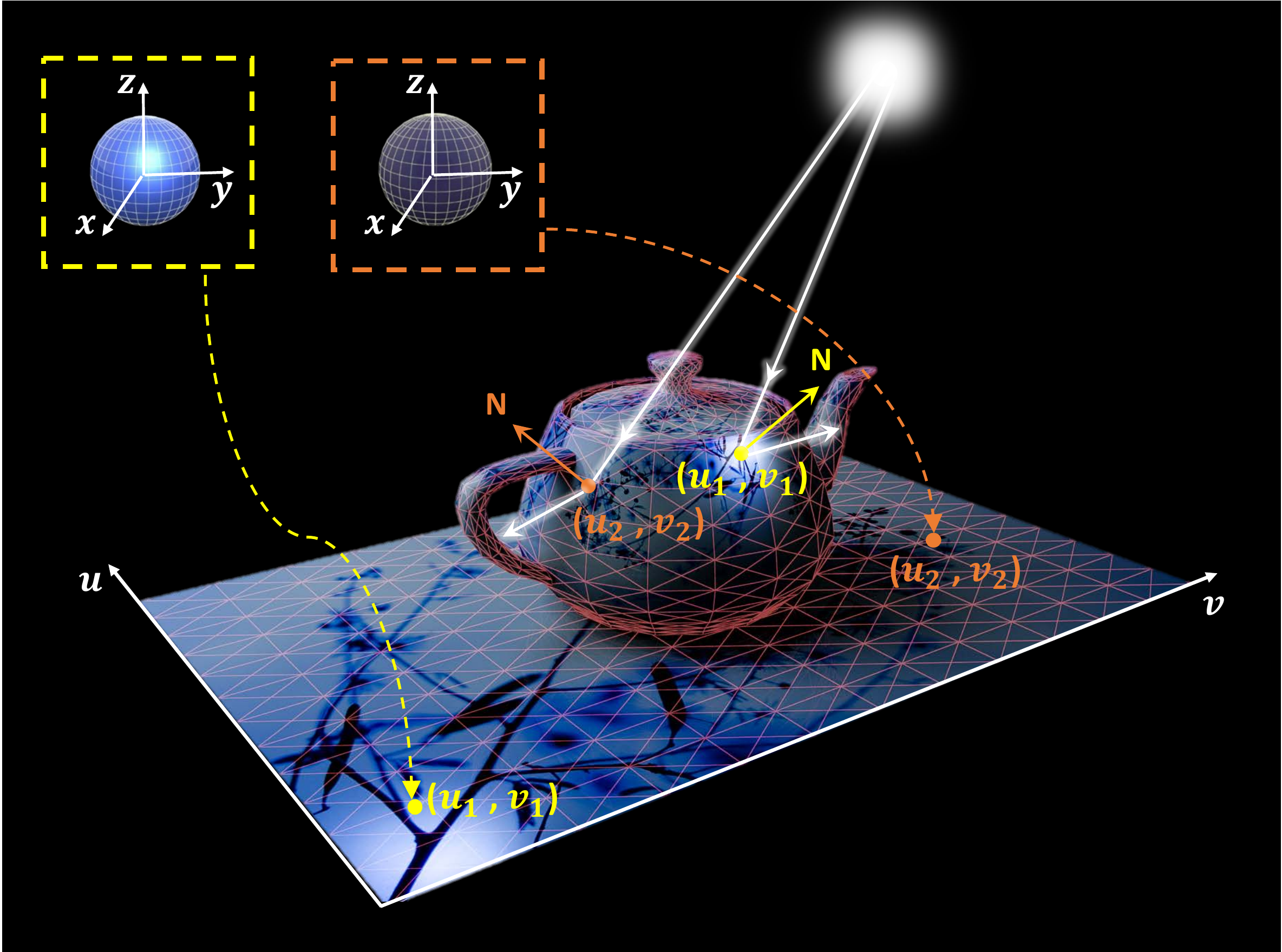}
\end{center}
   \caption{Our surface light field representation. We use the texture coordinate $[u, v]$ to index a surface point (ray origin) and $[d_x,d_y,d_z]$ to index the ray direction. The ray directions are further transformed (Section 3.1) to better preserve view dependent features.}
\label{fig:representation}
\end{figure}

In this paper, we develop a novel IBMR technique for modeling the surface light field (S-LF)~\cite{Mcmillan1995Plenoptic} by exploiting latest advances on neural networks~\cite{M1998Artificial,ren2013global}. An S-LF represents the radiance of rays originating from any point on the surface in any directions (Fig.\ref{fig:representation}) and hence serves as an image-based representation of surface reflectance. In reality, one needs to capture a very dense set of images at different viewpoints for faithful reproduction. Fig.\ref{fig:pipeline} shows an example of Tang San Cai, a special type of Chinese pottery made of lead-glazed earthenware, that exhibits semi-glossy and semi-metallic appearance. To ensure high quality rendering, traditional techniques based on ray interpolation require ultra dense sampling to minimize aliasing. 

We observe redundancies embedded within the S-LF. Consider an object composed of a single material (reflectance). Since the normal and view directions vary across vertices, the captured rays can be "fused" into dense angular samples. We hence employ a parallel data stream neural network as the "composer" as shown in Fig.~\ref{fig:NetworkStructure}. We call our solution Deep Surface Light Field or DSLF. In essence, the neural network plays dual roles: 1) it automatically fills in the missing data by combining different sampling patterns across the vertices, and 2) it eliminates redundancies while preserving view-dependent characteristics via learning.

To use DSLF, we enforce our data representations be continuous. We therefore adopt a 5D S-LF representation $(u, v, d_x, d_y, d_z)$ where $(u, v)$ index the vertices (as the origin of the ray) and $(d_x, d_y, d_z)$ to the viewing angle (as the direction of the ray). We pre-partition the input data into diffuse and view-dependent components and feed the latter into two parallel sub-nets. We add a skip-layer~\cite{he2016deep} to combine the final prediction layer with lower layers across finer strides, to conduct prediction under the global structure. To further enhance the network's separability to ray directions, we add an additional layer to connect the direction stream $(d_x, d_y, d_z)$ with the position stream $(u, v)$. 

We validate our DSLF framework on both synthetic and real objects with various surface reflectance models and under complex lighting conditions. For real data, we combine traditional structure-from-motion (SfM) with Perspective-n-Point (PnP) techniques for reliable image registration on highly specular objects. We also conduct texture-aware remeshing for aligning texture edges with vertices to avoid blurring. Experiments show that DSLF not only preserves high visual fidelity but also manages to achieve very high data compression rates. Finally, we implement the network decoder on the GPU for real-time rendering.

\section{Surface Light Fields}
Image-based modeling and rendering (IBMR) essentially aims to model the plenoptic function~\cite{Debevec1998Image}. Earlier work by McMillan and Bishop~\cite{Mcmillan1995Plenoptic} used the 5D ray representation (stored as panoramas captured at different 3D locations) for view interpolation. Levoy and Hanranhan introduced two plane parametrization or 2PP to reduce the 5D representation to 4D: each ray is represented by its intersections with two parallel planes pre-defined in 3D space, $st$ as the first intersection and $uv$ as the second. The goal of LF, however, is mainly to rendering parallax when moving viewpoints in 3D space. Later approaches exploit multi-texture for blending a large number of images (rays) to render dynamic refocusing~\cite{Levoy1996Light}. 

To use IBMR to model surface reflectance, Wood et al. proposed the surface light field (S-LF)~\cite{wood2000surface}. Given a vertex location, e.g., a vertex index $T = [u, v]$ which also corresponds to a specific 3D point on the object surface, we assume there is a ray along the direction $D_i = [d_x^i, d_y^i, d_z^i]$ hitting at $T$ and its corresponding reflection ray leaves along $D_o$, the angle between $D_i$ and $D_o$ is $\theta$. So we have the following simplified BRDF function:
\begin{equation}
\begin{split}
S(T, D_{o})
=&\int_{\mathcal{s}^2}f(T,D_{o},D_{i})L_{i}(T,D_{i}) | cos\theta | d D_{i}\\
\end{split}
\end{equation}
where $S(T,D_o)$ is the radiation of the ray from $T$ along direction $D_o$ in the upper hemisphere $\mathcal{s}^2$, $f(T,D_{o},D_{i})$ corresponds to the bidirectional transmittance distribution function (BTDF), and $L_{i}(T,D_{i})$ is the incoming radiance of point $T$ along direction $D_{i}$.

Using an S-LF for rendering, however, has been a dilemma. On one hand, sparsely sampled S-LF seems insufficient for view interpolation. On the other, a densely sampled S-LF can be overwhelmingly large. For example, sampling half a million surface points (e.g. total number of $T$) and five hundred different view-in directions $D_i$ produces over 10GB data. 

By far, most existing approaches have adopted "more is better" approach, i.e., they uniformly prefer dense sampling and then resort to compression schemes such as vector quantization (VQ)~\cite{Levoy1996Light}. Magnor and Girod~\cite{Magnor1999Adaptive} extended VQ via block-coding and 
on-line dequantization at the rendering stage. Miller et al.~\cite{Miller1998Lazy} developed a JPEG-like image compression scheme to only store the angular variations. Wood et al.~\cite{wood2000surface} combined generalized VQ and principal component analysis (PCA) by first constructing a codebook composed of a collection of prototype lumispheres and then computing their weights for individual patches. The decoding process of all these approaches is uniformly conducted on the CPU and does not ensure real-time performance.

A notable variation is the learning-based compression scheme~\cite{miandji2013learning} that exploits spatial coherence of the data via non-local clustering. Specifically, their approach represents the radiance distribution at every point  in terms of the hemispherical radiance distribution function (HRDF). They then cluster points with similar HRDF and learn a per-cluster compact dictionary. A major limitation is that their scheme requires nearly uniform sampling in the angular dimension, which is difficult to achieve in practice. 

Our approach also uses $S(T, D_o)$ to represent the complete surface light field of an object. Rather, we adopt the latest deep learning schemes that does not require uniform sampling. It is important to note that we use 3D vectors $(d_x, d_y, d_z)$ instead of the spherical coordinates $(\theta,\phi)$ to represent ray direction. This is because to train a deep network, it is essential that the parameters are continuous whereas $(\theta,\phi)$ exhibits discontinuous warping artifacts. Another major advantage of our DSLF is efficient rendering: when rendering a new viewpoint, we only just need to query the position and direction of each ray into the network. In Sec.\ref{sec:render}, we show querying the deep network for rendering can be directly mapped onto the GPU to achieve real-time performance.  

\section{DSLF Network}
We adopt a neural network structure resembling the autoencoder, a feedforward and non-recurrent neural network. Different from the traditional autoencoder where the output layer need to have the same number of nodes as the input layer for the purpose of reconstructing its own inputs, our DSLF, although having a similar structure, maps the 5D ray parameterization to 3D RGB color.  Our network learns the variation of lights in this environment and  material.
%
%
%
%

\subsection{Data Pre-procssing}
%
%
\label{sec:preprocess}

\begin{figure}[t]
\begin{center}
   \includegraphics[width=1\linewidth]{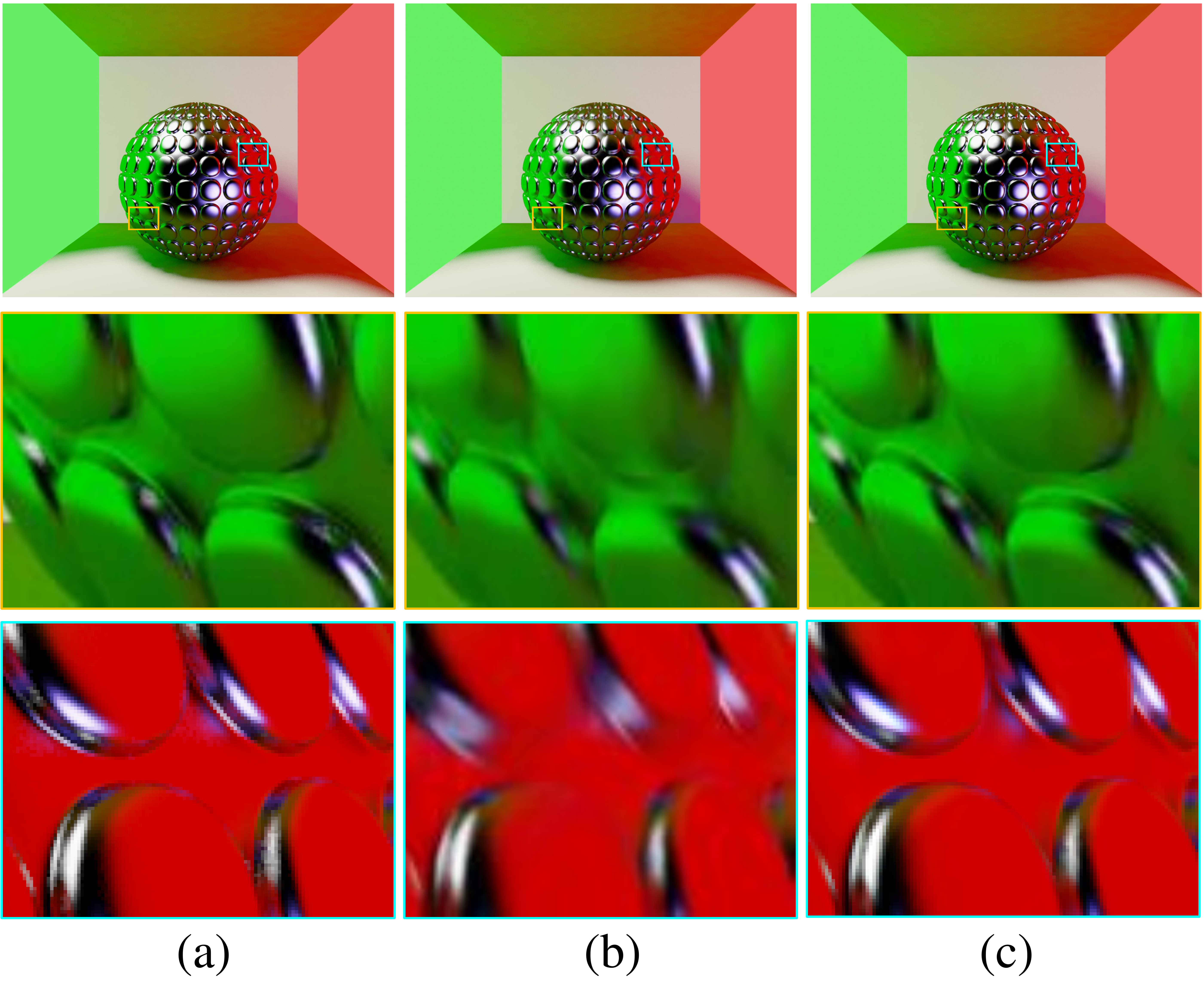}
\end{center}
   \caption{Results with and without data pre-processing. From left to right: (a) The ground truth. (b) Results without pre-processing exhibit excessive blurs. (C) Results with pre-processing preserves sharp edges and specularity.}
\label{fig:pre-processing}
\end{figure}

The brute-force approach would be to directly use the captured S-LF as inputs to the network. However, the raw S-LF consists of both low frequency (e.g. diffuse) and high frequency (e.g., specular) components where the latter is most important for realistic rendering. Training on the mixture will produce excessive blurs as shown in Fig.\ref{fig:pre-processing}. We therefore first separate the S-LF of a point $T$, $S(T,D_o)$ on the object into a diffuse component and a view-dependent (specular) component as:

\begin{equation}
\begin{split}
S(T,D_o)
&\approx \Phi_{d}(T) + \Phi_{r}(T,D_{o})
\end{split}
\end{equation}
where $\Phi_{d}(T)$ and  $\Phi_{r}(T,D_{o})$ are the diffuse and specular components individually.
For $\Phi_{d}$, we simply approximate it as the median color of the rays originating from $T$. Subsequently the specular component corresponds to the residue. We further convert the original lumispheres into the residual lumispheres in terms of the specular components. It is important to note that the residual lumispheres $\Phi_{r}(T,D_{o})$ should be generally sparse.

Consider two points $T_1 = [u_1,v_1]$ and $T_2 = [u_2,v_2]$ on the surface as shown in Fig. \ref{fig:representation}, illuminated by a point light source. The view direction that will observe strongest specularity corresponds to the reflected direction from the incident lighting direction. Conceptually, the reflected directions at $T_1$ and $T_2$ should observe similar specularity. However, due to normal differences, the two reflected directions (or $D_{o}$) in the S-LF representation are quite different. This imposes difficulty in training a neural network: the input directions are very different but the observed values are very similar.  
%
%
%
%
%
%

To resolve this problem, we "invert" the viewing direction $D_{o}$: for each view direction, we use the surface normal to inversely reflect the ray, e.t.  approximate lighting direction around the perfect specular reflection, $\tilde{D}_o$:
\begin{equation}
\tilde{D}_o = 2(\mathbf{n}^T \cdot D_o )\mathbf{n} - D_o
\end{equation}
where $\mathbf{n}$ is the normal at the point. For the rest of the paper, we assume the original S-LF $S(T,D_o)$ has been transformed to $S(T,\tilde{D}_o)$. 

Another key issue is to handle occlusions. At a surface point $T$, certain directions may be occluded due to self-occlusions. This can be easily detected, e.g., via z-buffer, if the geometry is known. In our solution, we simply ignore these rays, i.e., we do not use these rays in the training process. This should not cause much artifacts since the occluded point will not be observed in any case. The only case it will cause visual artifacts is when viewing the object at a really close distance, i.e., the captured views will not observe the occluded points but the virtual views might. 

\begin{figure*}[t]
\begin{center}
\includegraphics[width=1\linewidth]{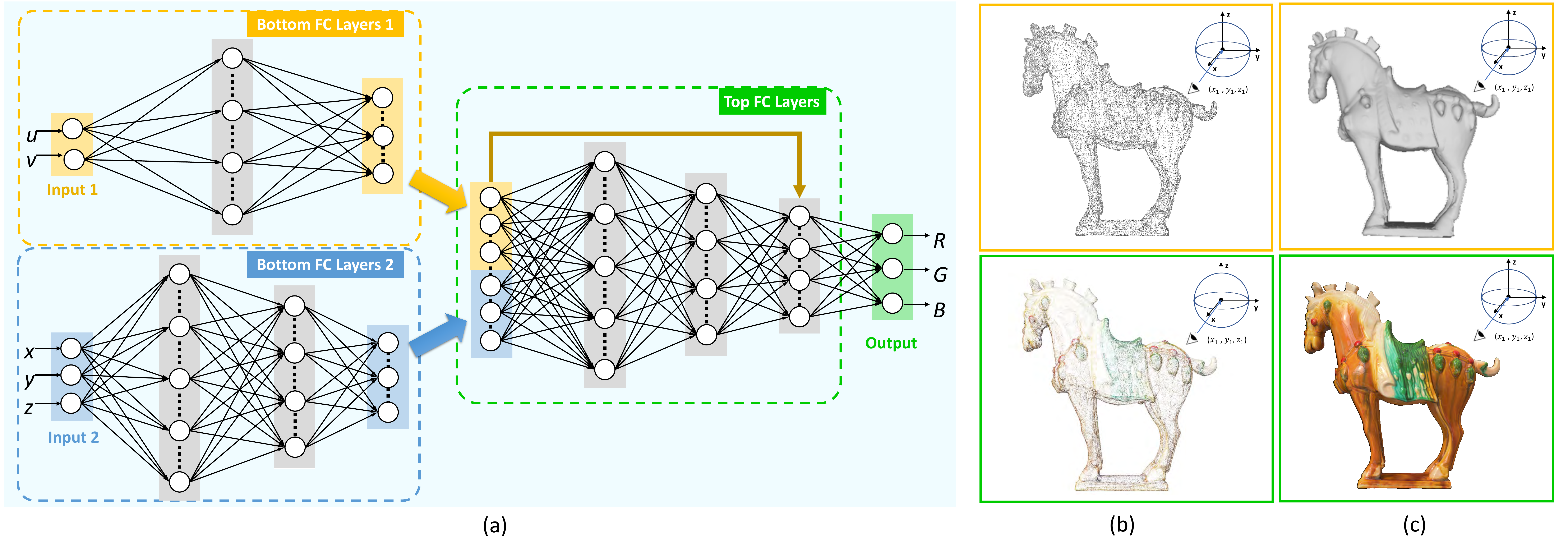}
\end{center}
   \caption{(a) Our deep surface light field (DSLF) network structure. The fully connected (FC) layers $\mathbf{L_1}$ and $\mathbf{L_2}$ subnets take ray coordinates as inputs and output to the FC $\mathbf{T}$ subnet, with an additional skip layer (yellow arrow). (b) shows a sample input(top) and per-vertex prediction(bottom). (c) shows the final output after rasterisation.}
\label{fig:NetworkStructure}
\end{figure*}

\subsection{Network Architecture}
Next, we compose a network to train our "inverted" 5D S-LF data. Recent advances on neural networks have demonstrated great success for conducting challenging reconstruction and scene understanding tasks with an end-to-end learning. Most recently, learning-based approaches have been adopted to rendering by exploiting unique characteristics of visual data \cite{radford2015unsupervised,miandji2013learning,ren2015image}. A specific goal there is to approximate continuous or square-integrable functions in rendering (e.g., BRDF, indirected illumination) via a feed-forward network \cite{hornik1989multilayer}. Specifically, inspired by \cite{chen2002light}, the S-LF reflection function can be approximated as a sum of a small number of products of lower-dimensional functions. As shown in Eq.\ref{eq:ref}, we use $g_k(T)$ and $h_k(D_i)$ to represent the reflection function $\Phi_{r}(T,D_{i})$.
%
%
%
%

\begin{equation}
\label{eq:ref}
\Phi_{r}(T,D_{i}) \approx \sum_{k=1}^K g_k(T)h_k(D_i)
\end{equation}

Further, continuous functions or square-integrable functions on finite domains can be approximated with arbitrarily small error by a feed-forward network with one or two hidden layers\cite{hornik1989multilayer}. Similarly, we use a fully-connected neural network to represent this function, as shown in Fig.\ref{fig:NetworkStructure}. 

Motivated by Eq.\ref{eq:ref}, we build the network with two parallel data streams: $\mathbf{L_1}$ and $\mathbf{L_2}$ who is fed with ray direction$(d_x,d_y,d_z)$  and vertex texture position $(u,v)$ respectively. An extra layer is added to represent object's surface, as we find the lighting distribution on surface space has higher dimension than viewing space, especially in complex ambient light conditions, and the representation ability of the model to light source most affects the sharpness of light edge (as shown in Fig.\ref{fig:node of network}).

In addition, we borrow the idea from the recent deep residual network~\cite{He2015Deep} and add one skip-layer to connect fine layers and coarse layers which enable the model to conduct local predictions constrained under the global structure. In our implementation, we adopt ReLU activation layers~\cite{Nair2010Rectified} due to its fast convergence rate and a sigmoid activation layer at the end since the residual contains negative value and needs to be clamped. 
%
%
%
%
%
%
%
%

The training procedure needs to update the unlinear mapping from the 5D S-LF parameter space onto the 3D color space with certain type of loss functions. However, traditional pixel-wise loss functions such as mean square error (MSE) fail to robustly handle uncertainty and lead to loss of high-frequency details such as textures. This is because minimizing MSE corresponds to computing averaging the pixels and therefore leads to over-smoothing. We instead adopt the Kullback-Leibler divergence loss function\cite{kullback1951information} on the residual RGB output to preserve the distribution rather than the mean value. 
%
%
%
%

Recall that the network should be able to encode the lighting information. We therefore use a large number of nodes. Subnet $\mathbf{T}$ retrieves information only related to  $\mathbf{L_1}$ ray directions and vertex positions (ray origins)$\mathbf{L_2}$. After we conduct extraction in the first stage, we feed the results to the second stage for conducting residual RGB inference. In our implementation, the fully connected nodes in $\mathbf{L_1}$ is $\{512,256\}$, and in $\mathbf{L_2}$ is $\{512,256,192\}$. The total number of nodes corresponding to different layers in network $\mathbf{T}$ is $\{1000,800,600,3\}$ individually. 

\section{Data Acquisition and Alignment}

Most previous S-LFs were obtained through rendering or using expensive capture setups. We aim to handle unstructured S-LF, e.g., the ones captured by a non-uniform rig or simply by moving a hand-held camera. Specifically, we describe the key steps on registering the sampled images onto the object model. The process is crucial as slight misalignment can lead to strong artifacts such as ghosting and blurring. We hence present a two-stage registration process.

\subsection{Geometry Acquisition and Image Registration}
\label{sec:registration}

\begin{figure}[t]
\begin{center}
   \includegraphics[width=1\linewidth]{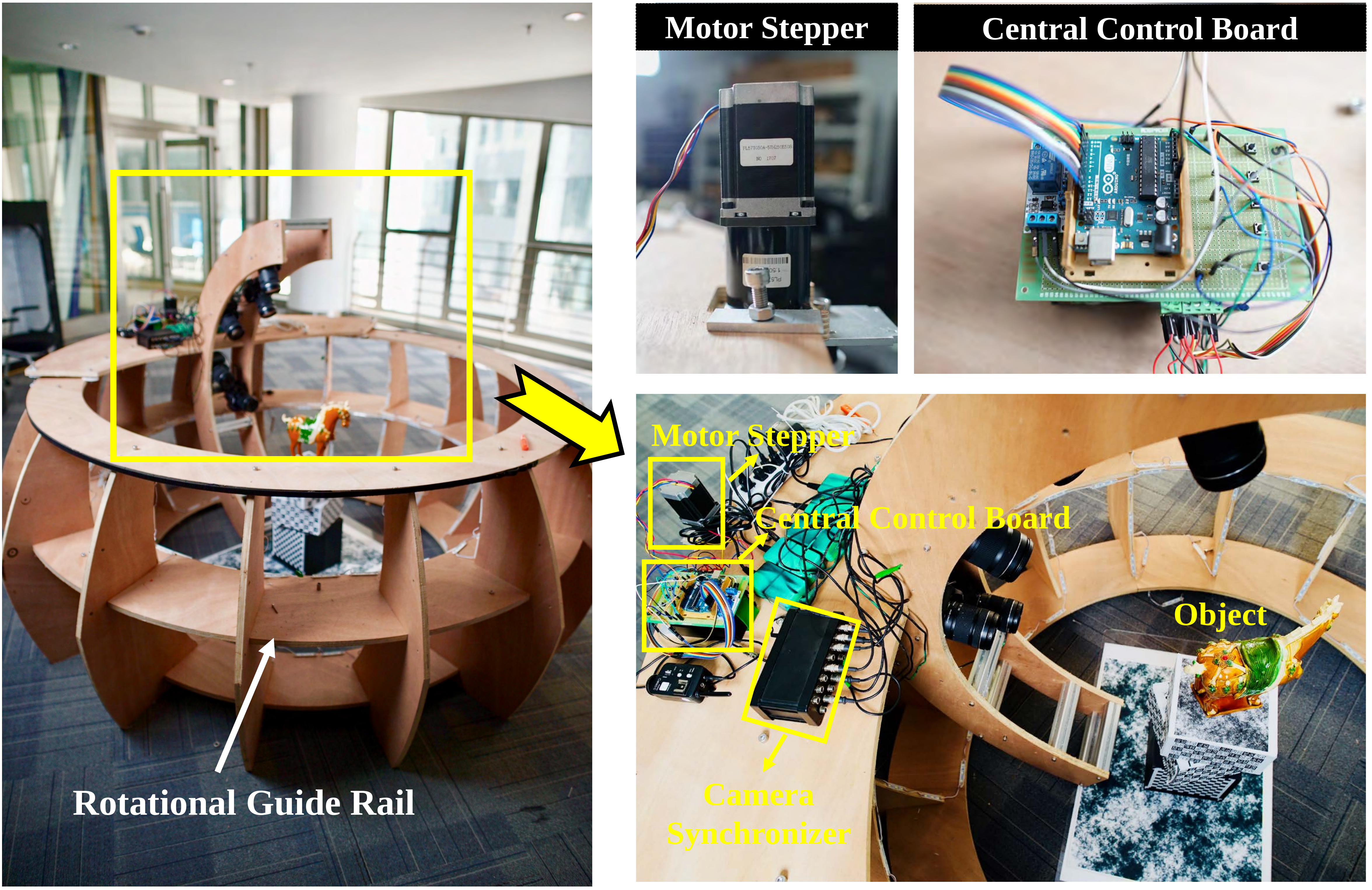}
\end{center}
   \caption{Our surface light field capture device mounts 5 cameras on a rotational arm that is controlled by a motor.}
\label{fig:device}
\end{figure}

In the first stage, we conduct 3D object reconstruction using computer vision techniques. It is paramount that we minimize view dependency effects to ensure robustness. Therefore, we first capture a sequence of images $\{I_{1}^{u}, I_{2}^{u}, ... I_{n}^{u}\}$ of an object
under nearly ambient illumination conditions (e.g., inside a soft light condition as shown in Fig.\ref{fig:device}). Each view should cover nearly the complete object so as to conduct
reliable feature matching. We adopt a modified version of Structure-from-Motion (SfM). 

We first conduct scale-invariant feature extraction and matching (SIFT/SURF) in each image and obtain image pairs $\{(I^u_i, I^u_j) | i, j \in n \}$ with a set of matched feature correspondences $M_{ij}$. Next, we use geometric verification and RANSAC~\cite{Martin1987Random} to reject the outliers: we assume a 3D point $X \in \mathbb{R}^3$ is projected to $I^u_i$ as $x_{i} \in \mathbb{R}^2$ and to $I^u_j$ as $x_{j} \in \mathbb{R}^2$; Notice that $x_{i}, x_{j}$ are subject to the epipolar constraint with the essential matrix $E$ for calibrated cameras or the fundamental matrix $F$ for uncalibrated ones. Further, if $X$ lies on a plane, we have $x_{j} = Hx_{i}$. We set out to find $H, E, F$ that can fit a sufficient number of feature points. If no such matrices exist, we deem the features as outliers. Finally, we conduct pose estimation using the verified feature correspondences. Specifically, we select two views with the maximum number of inliers as the initial image pair as $I^u_0, I^u_1$. For feature correspondences ${x_{i}, x_{j}} \in M_{01} $ between $I^u_0,I^u_1$, we have:

\begin{equation}
\hat{x_{j}}^{T} K_1^{-T} \hat{T}RK_0^{-1} \hat{x_{i}} = 0
\end{equation}
where $R \in \mathbb{R}^{3\times 3} ,T \in \mathbb{R}^3$ are the relative transformation between $I^u_0, I^u_1$ and $\hat{T}$ is the corresponding skew symmetry matrix of $T$. $\hat{x_{i}},\hat{x_{j}}$ represent the homogeneous representation of $x_{i},x_{j}$. $K_0, K_1$ are the intrinsic matrices of $I^u_0,I^u_1$ respectively. 

Notice that we can obtain $K_0, K_1$ directly from the captured image (generally provided by the camera vendor in the exif format). Therefore, we can directly use feature correspondences to solve $R,T$~\cite{Ma2004An} and then apply triangulation followed by bundle adjustment to estimate and refine both the 3D points and the poses. Next, we use the perspective-n-point (PnP) technique~\cite{Lepetit2009EPnP} to estimate poses of the rest views. The PnP method leverages 2D-3D correspondences and is very efficient and reliable. For instance, given three pairs of 2D-3D correspondences $A\leftrightarrow a, B\leftrightarrow b, C\leftrightarrow c$, where $A,B,C \in \mathbb{R}^3$ are the 3D points in world coordinate, $a,b,c \in \mathbb{R}^2$ are their corresponding 2D projected pixel locations in image $I$, we suppose $P$ is the camera optical center. So we have three distances and three angles, namely $PA, PB, PC, \alpha = \angle APB, \beta = \angle APC, \phi = \angle BPC$. According to law of cosines, we can get the following equation:

\begin{equation}
\begin{cases}
\begin{aligned}
PB^2 + PC^2 - 2~PB~PC~cos\phi - BC^2 = 0\\
PA^2 + PB^2 - 2~PA~PB~cos\alpha - AB^2 = 0\\
PA^2 + PC^2 - 2~PA~PC~cos\beta - AC^2 = 0
\end{aligned}
\end{cases}
\end{equation}

We can then solve for $PA, PB, PC$ and obtain the 3D coordinates $A_I, B_I, C_I$ of $A, B, C$ in current camera coordinates \cite{Gao2003Parallel}. This allows us to solve for $R, T$ of the new view and similarly we apply triangulation to obtain more 3D points. We iterate this process until all images are inserted and apply one more round of bundle adjustment:
\begin{equation}
E = \sum_{i}\sum_{j} w_{ij}(\Arrowvert P(I^u_j, X_i) - x_{ij} \Arrowvert ^2_2)
\end{equation}
where $I^u_j$ is correspondent to the $j$-th image, $X_i$ is the $i$-th 3D point in the space and $P(I^u_j, X_i)$ corresponds to the predicted 2D projection point of $X_i$ on the image $I^u_j$. $x_{ij}$ is the captured 2D projection point on the image $I^u_j$ for $X_i$. $w_{ij}$ here is a binary variable that denotes whether the 3D
point $X_i$ is visible in $I^u_j$. We use the Levenberg-Marquardt technique~\cite{Mor1977The} for optimization. Notice that depending on the features of the object, our modified SfM technique may generate a sparse 3D point cloud of the model. If the points are too sparse, they are unreliable for S-LF approximation. We use the estimated camera poses and parameters to conduct unstructured multi-view stereo~\cite{Sch2016Pixelwise} to obtain much denser geometry. 

In the second step, we capture another set of images under non-uniform lighting as the input S-LF $\{I_{1}^{s}, I_{2}^{s}, ... I_{m}^{s}\}$, e.g., by positioning the object under natural lighting and capturing it using a hand-held camera. Conceptually, one can repeat the process above to recover the geometry and then apply model registration such as the Iterative Closest Point (ICP) for registering the models and hence images. In reality, since we are most interested in objects with complex surface reflectance, these objects exhibit strong view dependent effects and are arduous to reconstruct under natural illuminations where the results cause large misalignment. 

We hence refine the initial ICP registration by reusing the PnP algorithm. Once we obtain the initial pose estimation of the S-LF inputs, we render a set of images $\{I_{1}^{r}, I_{2}^{r}, ..., I_{m}^{r}\}$ using the model and texture captured under ambient illumination at the same pose. We also render a set of companion depth image $\{D_{1}^{r}, D_{2}^{r}, ..., D_{m}^{r}\}$. Next we refine pose estimation using PnP: for each rendered image and acquired image, we conduct feature matching to obtain correspondences; since each feature point now also has depth information (from $D^r$), we can directly apply the PnP scheme to refine the estimation of the pose. In our experiments, the process is highly robust and reliable. 

\subsection{Texture-Aware Remeshing}

\begin{figure}[t]
\centering
   \includegraphics[width=1\linewidth]{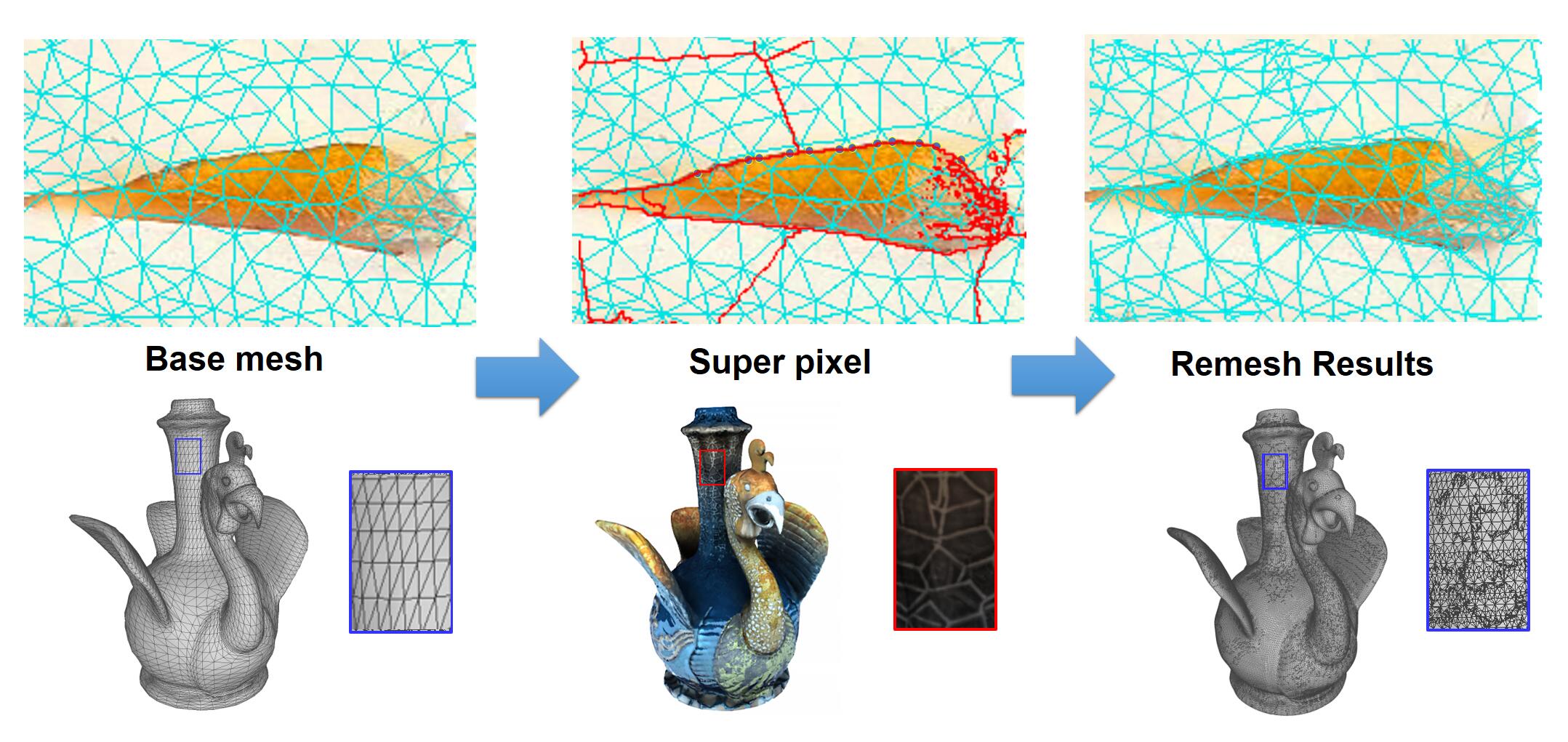}
   \caption{ 
   Our texture-aware remesh scheme. From left to right: part of the the original mesh and texture map, its superpixel segmentation, and the final remeshing result based on superpixels.}
\label{fig:remesh}
\end{figure}

Recall that SfM only estimates the 3D positions of feature points. If it misses a feature, it misses a vertex in the final mesh. Now, imagine a line feature on the texture of the object. Since SfM is only able to detect corners, it will miss all points lying on the line. Consequently, the triangulation process will produce a large, single triangle that contains the line. If we treat the triangle as a single entity, our deep network will blur the line features due to its convolutional nature.  

To resolve above problem, we further conduct a texture-aware remeshing scheme. Wood et al.\cite{wood2000surface} employed four-to-one subdivision connectivity method \cite{Lounsbery1997Multersolution}. However, this lazy wavelet-based process is complex and can generate unnecessary dense meshes that cause additional overhead to the deep network. We instead resort to the superpixel solution ~\cite{van2012seeds} that partitions pixels within each triangle into superpixels and then use them to guide the remeshing process.  

Superpixel segmentation is commonly treated as an energy minimization/maximization problem where each superpixel is defined as a region with a color distribution and a boundary~\cite{van2012seeds}. Our approach resembles the gradient-based level-of-detail method~\cite{Hu2010Parallel} that generates triangles only when needed (e.g., at texture edges). This significantly reduces the number of triangles. 

Recall that the interpolation function within a triangle is monotonic and smooth. Therefore we enforce the texture map within each triangle also be monotonic (i.e., the gradients are consistently positive or negative) and smooth (the second order derivative is nearly zero). Different from~\cite{van2012seeds} that computes the color likelihood, we evaluate the quality of a specific $k$ partitions in terms of its gradient consistency $C$, smoothness $G$, and shape priori $B$. Energy function $E$ could be presented as,

\begin{eqnarray}
E=C+\gamma G+\beta B
\end{eqnarray}
where $\beta$ and $\gamma$ weight the influence of each term and
\begin{eqnarray}
C= (\frac{1}{Z} \sum_{k}\sum_{\mathcal{A}}{sign(\Delta(i))})^2\\
G= \frac{1}{Z} \sum_{k}{\sum_{\lbrace H_{j} \rbrace} \sum_{i} \delta(\Phi(i)\in {H_{j}})}\\
B=  \sum_{i}\sum_{k}(\frac{1}{Z}\sum_{j \in \mathcal{N_{i}}}{\delta(j\in \mathcal{A})})^2
\end{eqnarray}
where $i$ is the pixel index within the patch, $\Delta(i)$ is the first order derivative at pixel $i$, $\Phi(i)$ is the second order derivative at $i$, $\delta$ is the step function, $Z$ is the normalization factor of the histogram. C measures the consistency of the sign of the gradients where the value is maximized when all $\Delta(i)$ are of the same sign. G measures the second order gradients to ensure smoothness. $\Phi(i)$ can be computed using the Laplacian operator. To improve robustness, we compute the histogram $H$ of $\Phi(i)$ and the term is maximized when $H$ is centered at zero and has a narrow spread. And $B$ is boundary smooth term and penalizes local irregularities boundaries and angles \cite{van2012seeds}, $\mathcal{A}$ contains the pixels in superpixel $k$, $\mathcal{N}_i$ be the path around pixel $i$.

\begin{figure}[t]
\centering
   \includegraphics[width=1\linewidth]{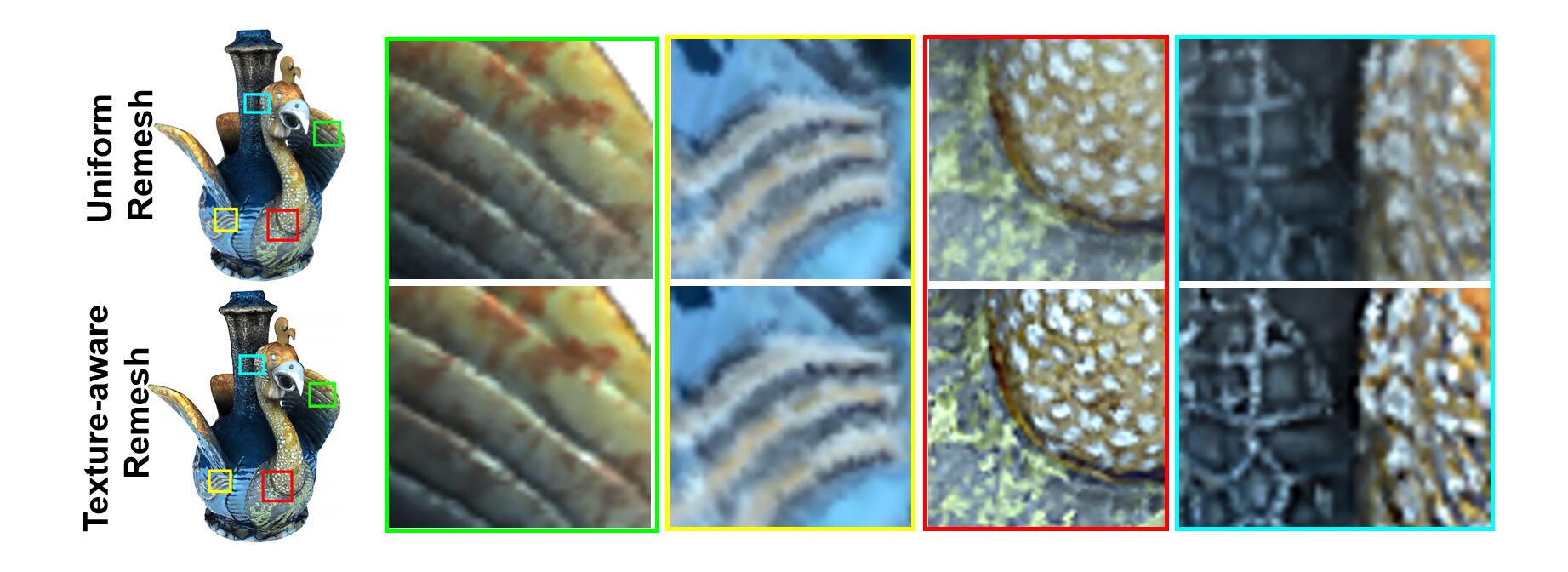}
   \caption{ 
   Rendering results based on uniform subdivision remeshing vs. our texture-aware remeshing. Our technique is able to preserve sharp edges and specularity.}
\label{fig:remesh2}
\end{figure}

To solve the optimal superpixel partitioning problem, we use Hill-Climbing optimization method to maximize energy $E(t)$. Fig.\ref{fig:remesh} demonstrates our remeshing process in detail. Given a basic mesh, we conduct the super pixel algorithm and then divide faces in the mesh along the edges. Fig.\ref{fig:remesh2} shows the rendering results before and after remeshing. The closeup images clearly indicate that the remeshing process makes the textures rendering more sharper. So our texture-aware remeshing scheme ensures the network encoded/decoded results to preserve sharp edges.  
>
\section{Experimental Results}
\label{sec:render}

We have experimented with our DLSF scheme on both synthetic and real data. 

\subsection{Results}
\begin{figure*}[t]
\begin{center}
   \includegraphics[width=0.88\linewidth]{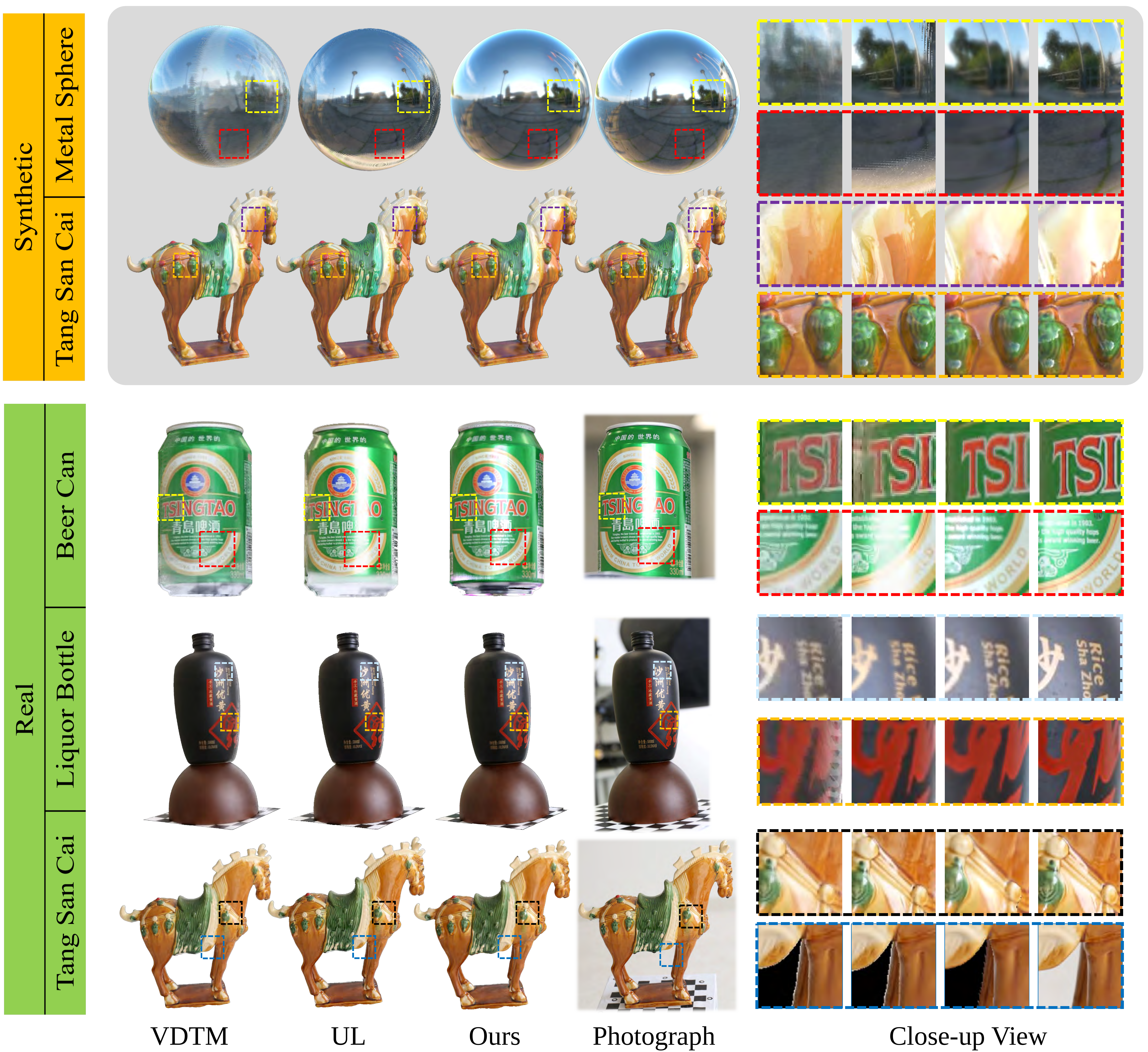}
\end{center}
    \caption{Comparisons of our DSLF vs. the state-of-the-art solutions view dependent texture mapping(VDTM) \cite{porquet2005real} and unstructured lumigraph (UL) \cite{Buehler2001Unstructured}.}
\label{fig:compare_synthetic}
\end{figure*}

For network training, we use the Adam \cite{kingma2014adam} optimization framework with $\beta_1 = 0.9$ and $\beta_2 = 0.999$ using an NVIDIA M4000 GPU. We employ a batch size of $1,500$ across all networks with $60,000$ iterations per epoch. We adopt a learning rate of $1e-4$ for the first $10$ epochs and reduce to $1e-5$ for the final $10$ epochs. 

\begin{figure*}[t]
   \includegraphics[width=0.8\linewidth]{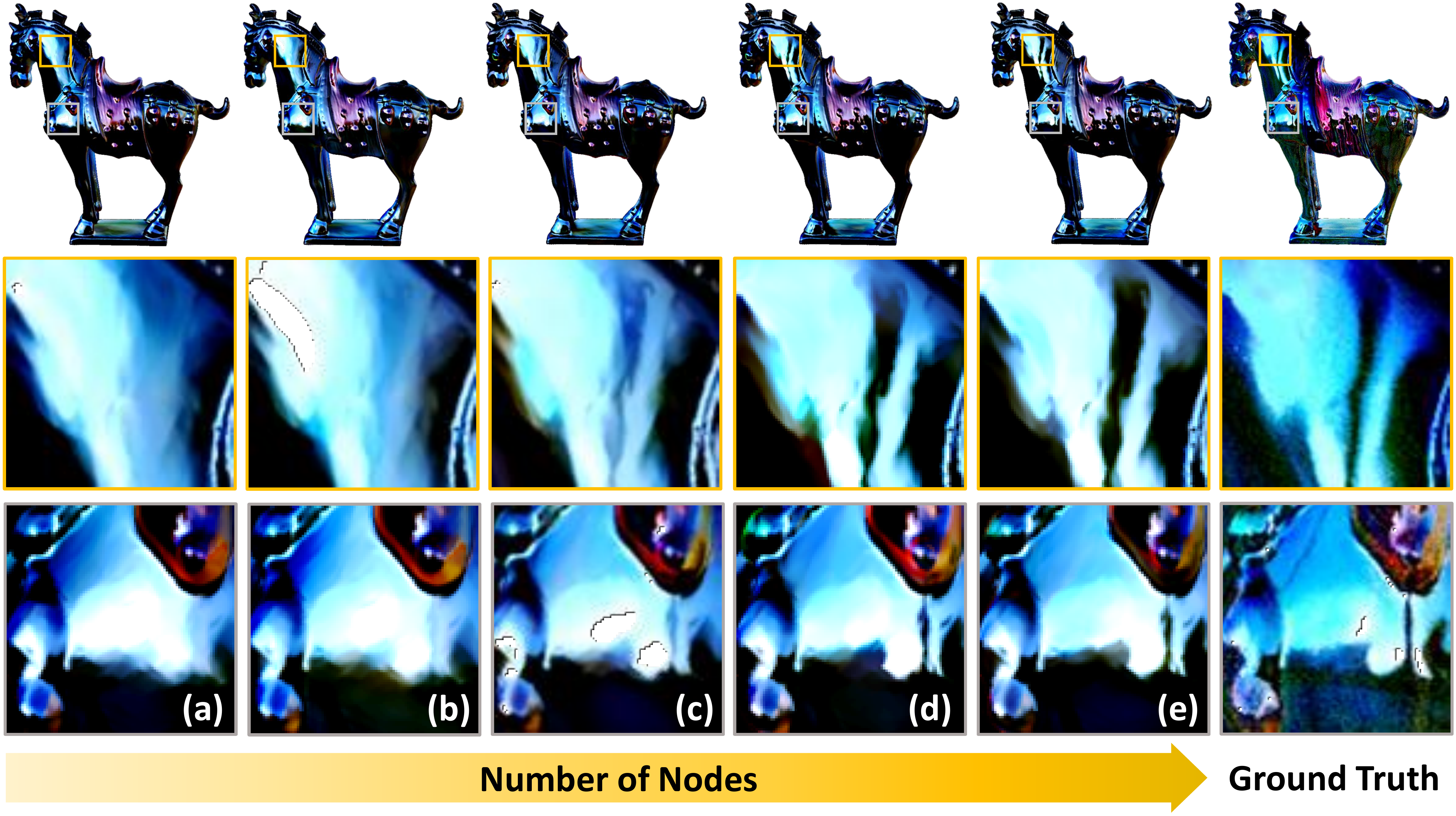}
   \caption{Rendering results with fixed $\mathbf{L_1}$ and $\mathbf{L_2}$ subnets but modifying the $\mathbf{T}$ subnet. (a) using only layer 1 and 4 in $\mathbf{T}$; (b) using layer 1, 2, 4 in $\mathbf{T}$; (c) using layer 1, 3, 4 in $\mathbf{T}$; (d) using all 4 layers; (e) adding an additional layer of 256 nodes between layer 1 and 2; (f) the ground truth.}
\label{fig:node of network}
\end{figure*}

On the Tang San Cai dataset, Fig.\ref{fig:node of network} demonstrates how different DSLF Net depths affect the rendering quality and the compression rate. We observe that initially as the network gets deeper, it produces better rendering quality, i.e., the results better preserve view-dependent features. Fig.\ref{fig:node of network} (e) and (f) illustrate that if we continue increasing the depth of the network, the rendering quality gets marginal improvements. However, the compression rate also drops accordingly. In our experiments, we found that using 3 hidden layers of top FC layers generally yields to satisfactory rendering quality and at the same time maintains a high compression rate as shown in Fig.\ref{fig:NetworkStructure}.

We implement DSLF by combining the CUDA and OpenGL pipelines to achieve real-time performance. Given a viewpoint and an object mesh (with normals and vertices), we compute, for each visible vertex $i$, the viewing direction $\mathbf{D}_i$ and then the "inverted" virtual viewing direction $\tilde{D_i}$ (Sec.\ref{sec:preprocess}) to obtain a 5D ray tuple. To index it into the trained network, we exploit efficient matrix computations on CUDA. Specifically, we stack all ray tuples into a matrix and store the DSLF Net also as a sequence of matrices. Since convolution is essentially matrix multiplications, we directly feed the ray matrix into the network and compute the per-vertex (ray) color via matrix multiplications. Recall every ray can be computed independently, we can further accelerate the process by assigning a thread to each ray. Finally, we use the standard rasterization pipeline to render each triangle from per vertex color. The complete algorithm is shown in Algorithm~\ref{al:render}.

\begin{algorithm}[t]
\caption{Rendering DSLF} 
\begin{algorithmic}[1]
\Procedure{$RenderOneFrame(\mathbf{T},\mathbf{N},\mathbf{F},\mathbf{C})$}{}
\State $(\mathbf{T'},\mathbf{N'},\mathbf{F'}) = BackFaceCulling(\mathbf{T},\mathbf{N},\mathbf{F},\mathbf{C})$
\For{each vertex $\mathbf{T}_i$ in $\mathbf{T'}$}
   \State $\mathbf{d}_i$ = CalculateViewingDirection($\mathbf{T}_i$,$\mathbf{C}$)
   \State $\tilde{\mathbf{d}_i}$ = InvertTransform($\mathbf{d}_i$,$\mathbf{n}_i$)
\EndFor
\State $\tilde{\mathbf{D}} = (\tilde{\mathbf{d}_1}, ..., \tilde{\mathbf{d}_i})$
\State VertexColor = DSLF($\mathbf{T'}$, $\mathbf{\tilde{\mathbf{D}}}$)
\State Image = ProjectAndRasterizate($\mathbf{C}$,$\mathbf{T'}$,$\mathbf{F'}$,VertexColor)
\State \Return Image
\EndProcedure
\end{algorithmic}
\label{al:render}
\end{algorithm}

\begin{figure*}[t]
\begin{center}
   \includegraphics[width=0.688\linewidth]{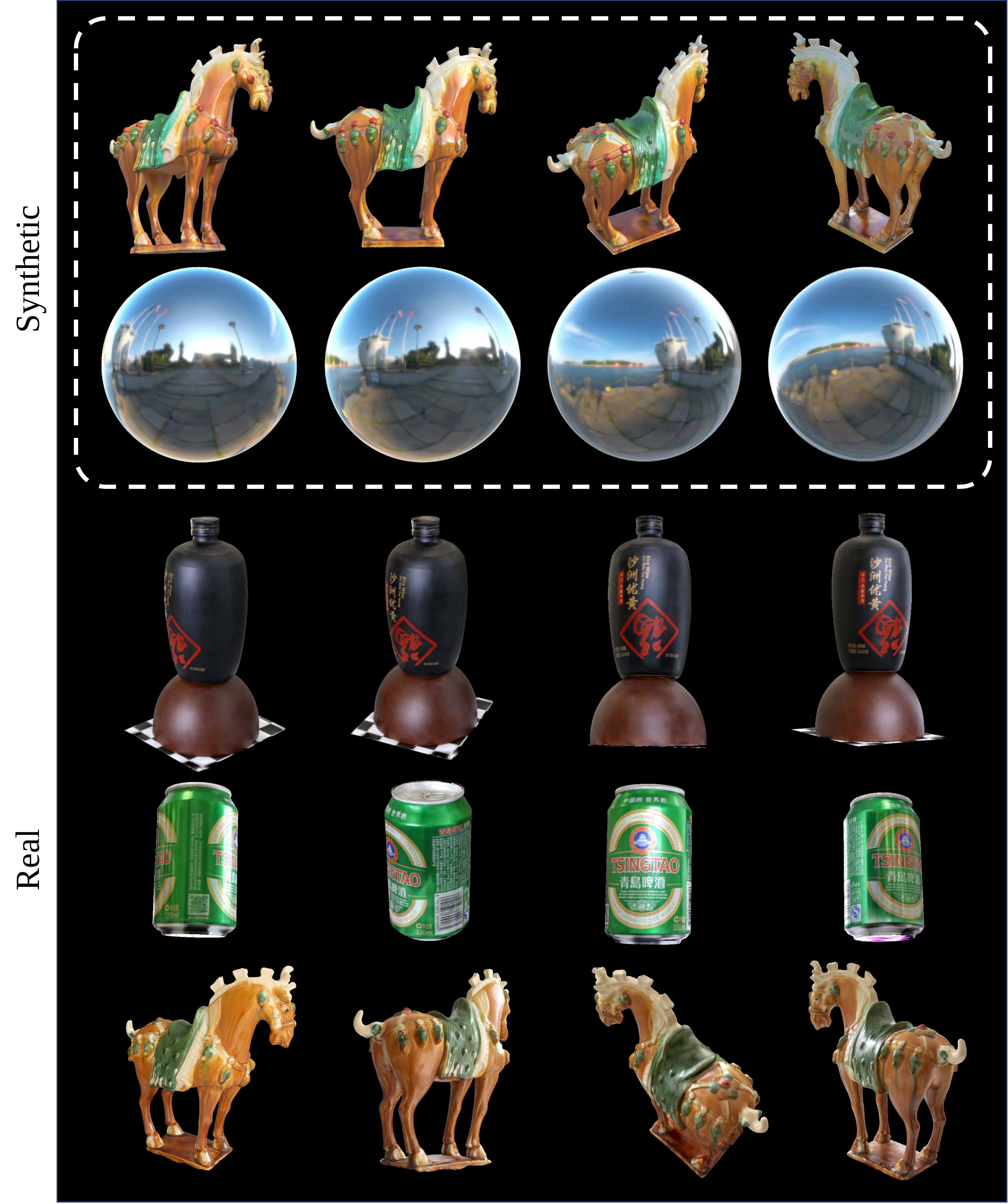}
\end{center}
   \caption{DSLF rendering results from different viewpoints . We show our highly photorealistic rendering appearance of objects with different reflection materials in both synthetic and real scene experiments: ceramics (top and bottom rows), metal (second and fourth rows), wooden (third row).}
\label{fig:result}
\end{figure*}

All experiments were conducted on an NVIDIA GeForce GTX 1080 graphics card with 8GB memory. We render S-LFs on objects composed of various materials and under different lighting conditions. For quantitative comparisons, we use the peak signal-to-noise ratio (PSNR) and the structural similarity index (SSIM) for measuring the rendered vs. ground truth data. We compare our technique with the state-of-art VDTM\cite{porquet2005real} and unstructured lumigraph\cite{Buehler2001Unstructured} methods in our implementation. 

\begin{table}
\begin{center}
\begin{tabular}{l|c|c|c|c|c} 
 \textbf{System Setup} & Image \#&Vertex \# & Superpixel  \#\\
  \hline
Metal Sphere & 200 &521,962& 6,000 \\
Synthetic Tang San Cai & 215&432,421 & 6,000 \\
Beer Can&320&467,418&4,000\\
Liquor Bottle&217&483,056&4,000\\
Real Tang San Cai &216&441,412&4,000\\
\end{tabular}
\end{center}
\caption{Specs of the input data for our experiments.}
\label{tb:system_setup}
\end{table}

\begin{table*}
\begin{center}
\begin{tabular}{l|c|c|c} 
\textbf{ } & DSLF & VDTM & UL \\
\hline
\quad Metal Sphere & 27.1997 \qquad0.9194 & 19.0778 \qquad0.8164 &21.6797 \qquad0.8339\\
\quad Synthetic Tang San Cai & 28.5230 \qquad0.9687 & 23.6546 \qquad0.9415&25.4801 \qquad0.9594\\
\quad Beer Can &18.0536\qquad 0.6555&13.0823\qquad 0.4851&14.5006\qquad 0.5741\\
 \quad Liquor Bottle &22.2871\qquad0.8099&20.8831\qquad0.7727&20.8893\qquad0.7912\\
\quad Real Tang San Cai &20.4035\qquad0.9480&21.4786\qquad0.9242&21.6177\qquad0.9350\\
\label{tb:psnr_ssmi}
\end{tabular}
\end{center}
\caption{Quantitative comparisons of DSLF vs. VDTM and UL. We use PSNR (left column) and SSIM (right column) as metrics. For each object, we render 20 random viewpoints at a resolution of 1000x1000 and only compute the errors on effective pixels (pixels that see the object).}
\label{tb:psnr_ssim}
\end{table*}

\paragraph{Synthetic Scenes.} For synthetic scenes, we use 3ds Max to render two datasets, e.g., a highly reflective metal ball and a Chinese pottery Tang San Cai. We position the object under natural lighting environments. For the sphere dataset, we use an initial mesh of 512,962 vertices and 1043,920 faces. We render $200$ images, each at a $1536 \times 1167$ resolution. The camera positions were randomly selected on a sphere surrounding the object.  Our remeshing scheme partitions the original mesh into $6,000$ superpixels. The Tang San Cai dataset uses $215$ images with the same resolution at the sphere. Our remeshing scheme produces $6,000$ superpixels (Table.\ref{tb:system_setup}).

Fig.\ref{fig:result} shows our DSLF rendering results. From different view directions, our algorithm is able to produce highly realistic virtual views, faithfully illustrating detailed lighting changes and material characteristics. The top two rows in Fig.\ref{fig:compare_synthetic} illustrate the close-up views of the rendering results using our method vs. VDTM~\cite{porquet2005real} and unstructured lumigraph (UL) ~\cite{Buehler2001Unstructured}. VDTM and UL rendering are unable to recover spatially sharp and angularly coherent specularities: direct interpolation not only produces polygonal specularity but also incurs excessive blurs. The artifacts are largely attributed to small spatial/angular support: they only use "local" information for view interpolation. Our DSLF, however, uses information learned from all angular samples and much better preserves specularity. In addition, our network effectively compresses the data, reducing the raw data size by multiple orders of magnitudes. 

\begin{table*}
\begin{tabular}{l|c|c|c} 
 \textbf{ } & Input Size (GB)   & Compression Rate & Rendering Speed (FPS) \\
  \hline
Metal Sphere & 2.003 &2567:1& 42 \\
Synthetic Tang San Cai &1.3&2760:1 & 41\\
Beer Can&4.89&6267:1&38\\
Liquor Bottle&3.36&4309:1&38\\
Real Tang San Cai &3.35&4289:1&35\\
\end{tabular}
\caption{Raw LF data size, DSLF LF data size (Compressed Output size is 0.79 MB), compression rate (Raw/DSLF), and rendering speed of DSLF implemented on an NVIDIA M4000 GPU. }
\label{tb:Compression}
\end{table*}

\paragraph{Real Scenes.} We use a Canon 760D camera to capture several real datasets including a real Tang San Cai, a beer Can and a ceramic liquor bottle. Table.\ref{tb:system_setup} shows the detailed parameters of the camera and the scene. Our two-stage registration scheme (Sec.\ref{sec:registration}) manages to align the models with their corresponding images even under highly complex lighting conditions. Fig.\ref{fig:compare_synthetic} shows our reconstruction results. 

Rendering objects such as beer cans and ceramic bottles is traditionally challenging for image-based rendering. Such materials are anisotropic and incoherent (silver logo vs. green paint in the beer can and white vs. red prints on the ceramic bottle) as shown in Fig.\ref{fig:compare_synthetic}. VDTM and UL generate new views in the image space and the interpolation scheme causes ghosting artifacts. In contrast, DSLF exploits dense angular samples across different locations over the surface and manages to render highly sharp images. Further, DSLF produces smoothly varying specularity across the views. We refer the readers to the supplementary video for the complete results.

Table.\ref{tb:Compression} shows that our network based approach can compress gigabyte (GB) raw input LF to under 1 megabyte (MB) DSLF, achieving a very high compression while still rendering at an interactive speed. Even at such a high compression rate, DSLF still maintains higher PSNR and SSIM metrics than the uncompressed VDTM and UL rendering, as shown Table.\ref{tb:psnr_ssim}.

\section{Conclusions}

We have presented a novel deep network based surface light field modeling and rendering technique called DSLF. At the core is the use of deep network's predictability and compression capabilities for handling view-dependent characteristics in a surface light field. We have further presented techniques for data acquisition and registration as well as mesh pre-processing and real-time rendering, to practically use our solution for real scenes.  

Our current solution exploits parallel processing on the GPU to conduct convolution operations in neural networks. Although we can achieve real-time performance even on low-end GPUs, the solution is not yet real-time on mobile devices. An immediate future task is hence to simplify the network and fit it on mobile GPUs. There is also an emerging trend on designing deep learning mobile chips and, if success, our solution can be directly mapped onto such architecture. Finally, we would like to investigate lighting separation and lighting transfer. In our model, we separate the specular highlights by assuming the median corresponds to the diffuse component. It will be highly useful to learn a more precise parametric model from the input data, possible by employing physically-based rendering engines, and then transfer lighting to different objects to achieve image-based relighting.



\bibliographystyle{ACM-Reference-Format}
\bibliography{sample-bibliography}

\end{document}